\documentclass[a4paper]{jpconf}
\usepackage{graphicx}

\usepackage{graphicx}
\usepackage{amssymb}
\usepackage{amsmath}    % need for subequations
\usepackage{txfonts}
\usepackage{xspace}

\newcommand{\He}{H.E.S.S.\xspace}

\newcommand{\M}{M\,87\xspace}

\begin{document}

\title{The \He extragalactic sky}

\author{Martin Raue}

\address{Institut f\"ur Experimentalphysik, Universit\"at Hamburg, Hamburg, Germany}

\ead{martin.raue@desy.de}

\author{for the \He Collaboration}

\begin{abstract}
The \He Cherenkov telescope array, located on the southern hemisphere in Namibia, studies very high energy (VHE; $E>100$\,GeV) $\gamma$-ray emission from astrophysical objects. During its successful operations since 2002 more than 80 galactic and extra-galactic $\gamma$-ray sources have been discovered. \He devotes over 400\,hours of observation time per year to the observation of extra-galactic sources resulting in the discovery of several new sources, mostly AGNs, and in exciting physics results e.g. the discovery of very rapid variability during extreme flux outbursts of PKS\,2155-304, stringent limits on the density of the extragalactic background light (EBL) in the near-infrared derived from the energy spectra of distant sources, or the discovery of short-term variability in the VHE emission from the radio galaxy M\,87. With the recent launch of the Fermi satellite in 2008 new insights into the physics of AGNs at GeV energies emerged, leading to the discovery of several new extragalactic VHE sources. Multi-wavelength observations prove to be a powerful tool to investigate the production mechanism for VHE emission in AGNs. Here, new results from \He observations of extragalactic sources will be presented and their implications for the physics of these sources will be discussed.
\end{abstract}

%------------------------------------------------------------------------------------------------------------------------
% Introduction
%------------------------------------------------------------------------------------------------------------------------
\section{Introduction}

\begin{figure}[tb]
\centering
\includegraphics[height=0.43\textwidth]{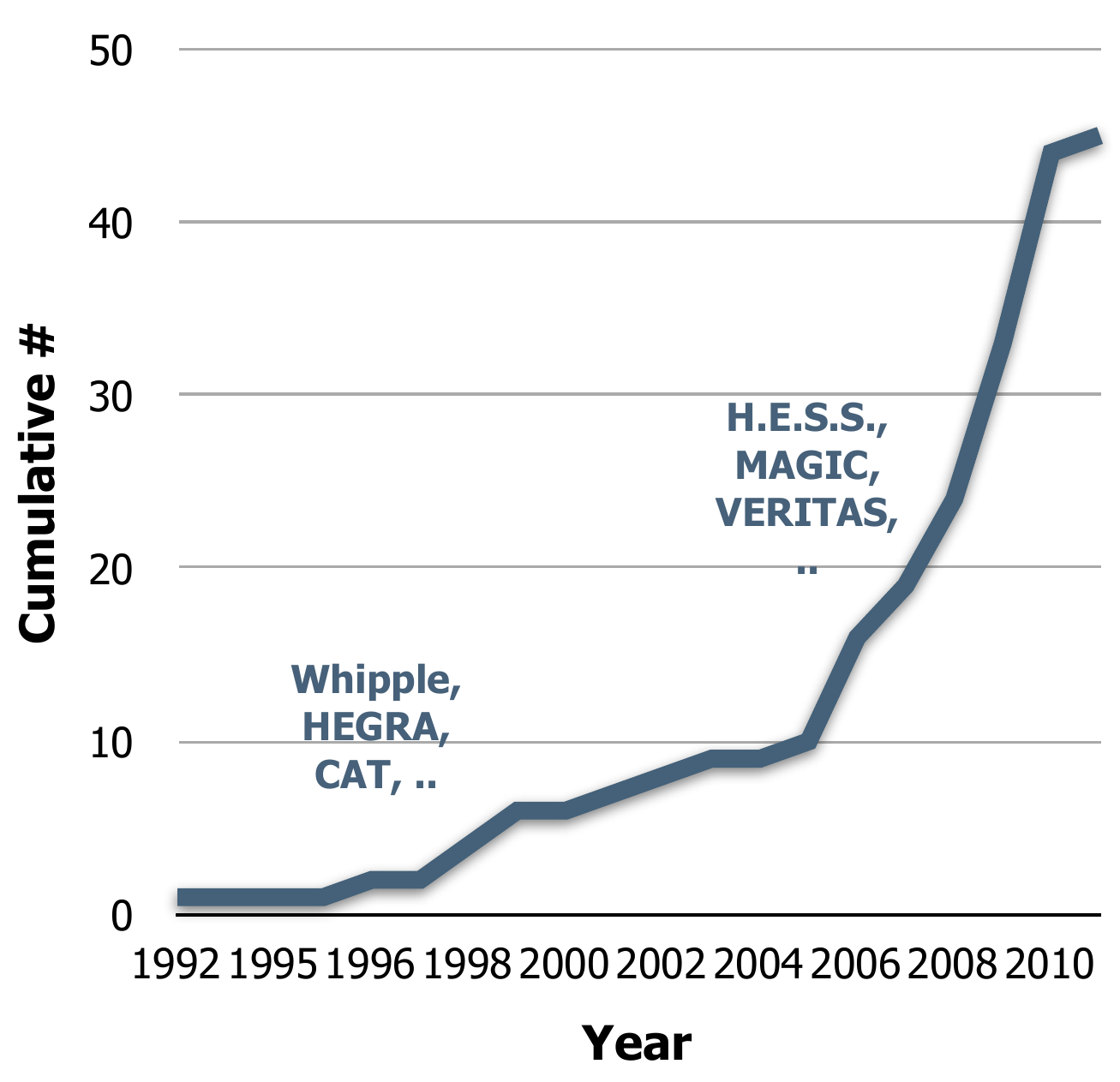}\hfill\includegraphics[height=0.42\textwidth]{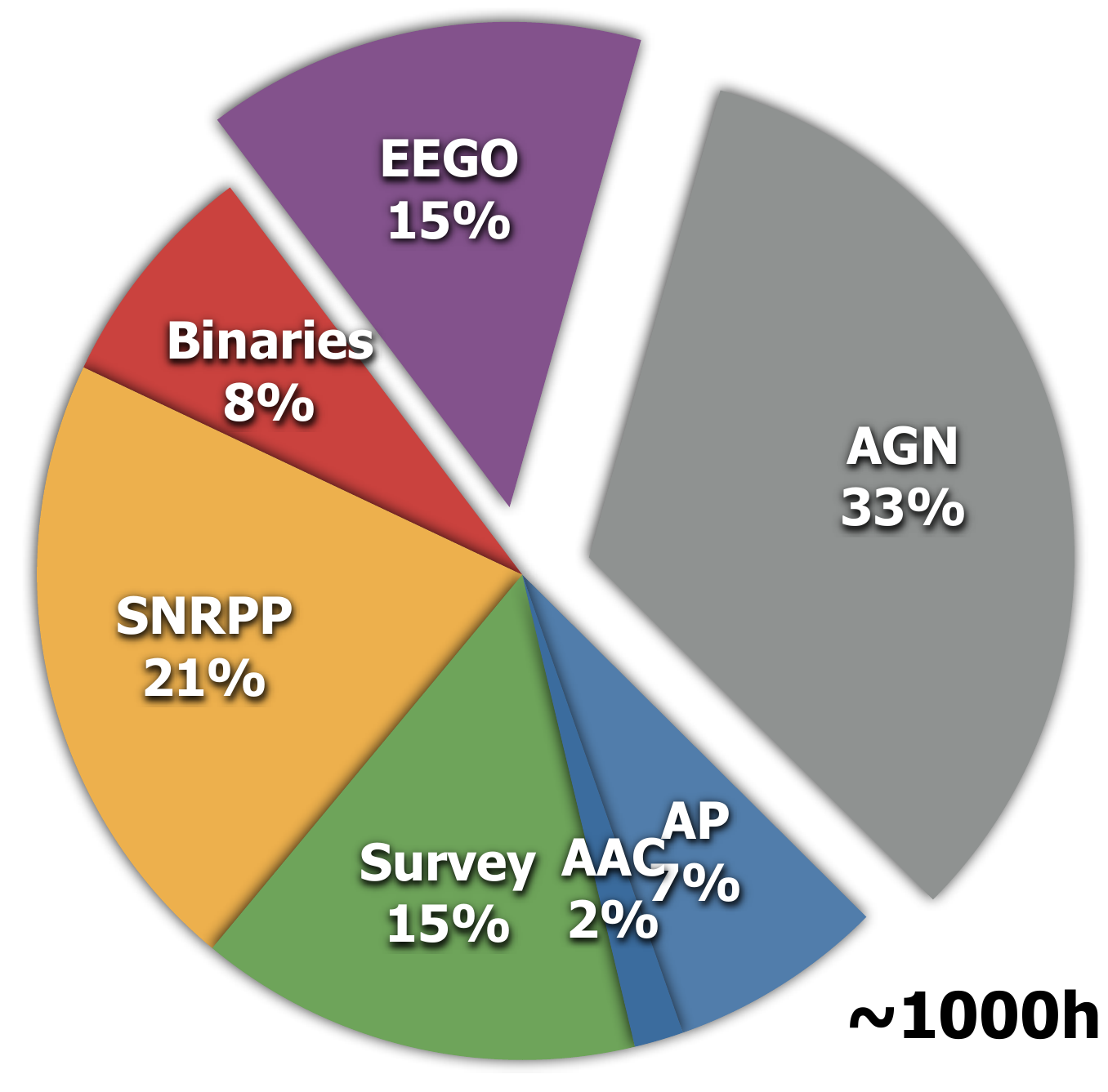}
\caption{\textit{Left:} Cumulative number of extragalactic VHE sources from TeVCat (\texttt{http://tevcat.uchicago.edu/}).\label{Fig:VHESources} \textit{Right:} Typical time allocation for different observing programs in H.E.S.S. EEGO: Extended extragalactic objects; AGN: Active galactic nuclei; AP: Astroparticle; AAC: Analysis and calibration; SNRPP: Supernova remnants, pulsars, and plerions. The total observing time per year is of the order of 1000\,h.\label{Fig:HESSObsTime}}
\end{figure}

Very high energy (VHE; $E>100$\,GeV) $\gamma$-ray astrophysics of the extragalactic sky is highly dynamic field. Since the detection of the first source in 1992 (Mkn\,421 by Whipple; \cite{punch:1992a}) the number of sources has been growing continuously to more than 45 known sources at the beginning of 2011 (see Fig.~\ref{Fig:VHESources} left). The detected source population is still dominated by active galactic nuclei (AGN), but recently starburst galaxies have also been established as extragalactic VHE emitters \cite{acciari:2009:veritas:m82:nature,acero:2009:hess:ngc253:science}.

The key science questions in the field are (still) investigating the origin and physics mechanism responsible for the VHE emission in AGNs and thereby probe the conditions of particle acceleration and cooling in relativistic plasma outflows and in the vicinity of super-massive black holes (SMBH).
One of the most striking observational result of the recent years was the discovery of rapid variability with minute time-scales \cite{aharonian:2007:hess:pks2155:bigflare,albert:2007:magic:mkn501}, which has not yet fully been integrated in the theories. Such short variability time-scales strongly impact the canonical modeling of the VHE emission in synchtrotron-self-Compton (SSC) frameworks, requiring large bulk Lorentz factors of the emission region ($\Gamma > 50$) \cite{begelmann:2008a}.
In addition, new source classes have been established as VHE emitters e.g. radio galaxies (M\,87, Centaurus\,A), low- and medium-frequency peaked BL Lac objects, etc.
With the launch of the Fermi satellite in 2008, a wealth of new information on the extragalactic sky in the 100\,MeV to 100\,GeV energy regime became available, strongly influencing the science programs in the VHE domain.

The multi-wavelength (MWL) aspect, in general, plays a crucial role in investigating VHE emission from AGN: The limited angular resolution of VHE instruments usually does not allow for a precise spatial determination of the origin of the emission in extragalactic objects. Investigating the MWL behavior e.g. by searching for correlations with high resolution radio observations allows one to probe spatial scales down to a few hundred Schwarzschild radii of the SMBH \cite{acciari:2009b:m87joinedcampaign:science}. New insights into the structure of the magnetic fields in AGN jets also come from polarimetric observations in the optical and radio.

Extragalactic VHE sources are also successfully used to study the meta-galactic radiation and magnetic fields (e.g. \cite{aharonian:2006:hess:ebl:nature,mazin:2007a,neronov:2010a}). In such studies, the extragalactic VHE sources are used as \textit{lighthouses}, whose emission gets altered by the intervening fields. For example, in case of the extragalactic background light (EBL), an attenuation of the VHE photons from pair-production with the low energy photons from the EBL is expected. With assumptions about the intrinsic spectrum emitted at the source constraints on the intervening fields can be derived (e.g. \cite{aharonian:2006:hess:ebl:nature,mazin:2007a}).

%------------------------------------------------------------------------------------------------------------------------
% H.E.S.S. status
%------------------------------------------------------------------------------------------------------------------------
\section{H.E.S.S. status \& extragalactic program}

% H.E.S.S. general
The High Energy Stereoscopic System (\He) is an array of four Imaging Atmospheric Cherenkov Telescopes (IACT) for the detection of very high energy (VHE; $E>100$\,GeV) $\gamma$-ray emission from astrophysical objects \cite{hinton:2004a:hess:status}. \He is located on the southern hemisphere in Namibia ($23^\circ16 ' 18 '' $S, $16^\circ30'00''$E) at an altitude of 1800\,m. It detects the Cherenkov light emitted by extensive air showers initiated by $\gamma$-rays or charged cosmic rays entering the Earth's atmosphere using ultra-fast photomultiplier cameras with 968 pixels located in the focal plane of 13\,m optical telescopes (tessellated mirror). The array is sensitive to $\gamma$-rays in the energy range from $\sim$100\,GeV to $\sim$100\,TeV with a typical energy resolution of $\lesssim15$\%. \He has been in operation since 2002, the full four telescope array since 2004. During its 9\,years of operations \He observations led to the discovery of over 80 new VHE sources and significant contribution to the field of VHE astrophysics and beyond.

\begin{figure}[tb]
\centering
\includegraphics[width=\textwidth]{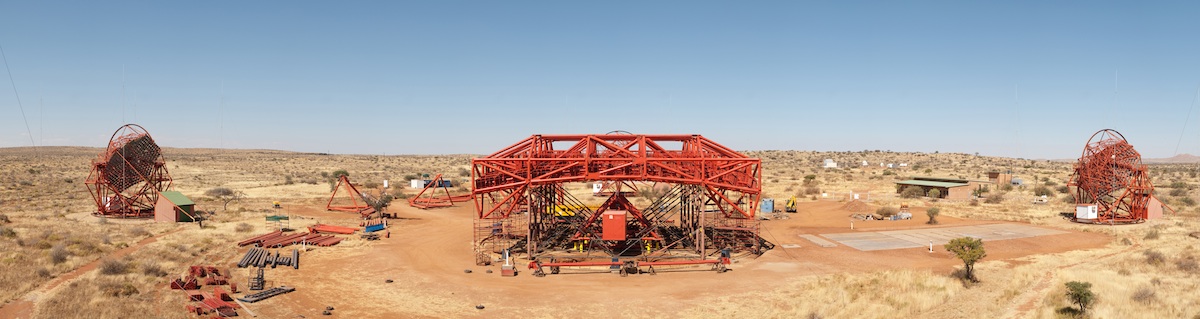}
\caption{\He site in January 2011. In the center of the image the dish support structure of the \He II telescope is visible, at the left and right edge two of the four \He I telescopes can be seen (Image credit: Julien Bolmont).\label{Fig:HESSSite}}
\end{figure}

% H.E.S.S. status / future
In 2010/11 the mirrors of the \He telescopes are being re-coated to lower the energy threshold back to the original value of $\sim100$\,GeV\footnote{The reflectivity of the mirrors decreases over time due to environmental effects (dust, wind, ..), which leads to an increased energy threshold. This is being accounted for in the analysis by evaluating the optical throughput of the system from the ring images of individual muons.}. At the same time, construction of the next extension of \He, a very large (25\,m) single telescope in the center of the array, is under way with first light expected in 2012 (see Fig.~\ref{Fig:HESSSite}).

% H.E.S.S. extragalactic program
Observing the extragalactic sky is one of the key science topics in \He Fig.~\ref{Fig:HESSObsTime} right shows the typical distribution of the \He observation time for one season (here 2011) on the different science groups, with the extragalactic topics taking about 50\% of the total time of $\sim$1000\,h. The extragalactic science in \He is divided into two sub-topics: (1) physics related to active galactic nuclei (AGN) and (2) extended extragalactic objects (EEGO), which includes galaxy clusters and starburst galaxies. The extragalactic program includes discovery observations, monitoring of known sources, multi wavelength campaigns (MWL), and deep studies of individual sources. In addition, extragalactic source are also used to study the intervening photon fields, mainly the extragalactic background light (EBL), and magnetic fields.

In the following, selected highlights from the past 2 years of \He observations of the extragalactic sky are presented.

%------------------------------------------------------------------------------------------------------------------------
% HESS\,J1943+213: a candidate extreme BL Lac object
%------------------------------------------------------------------------------------------------------------------------
\section{HESS\,J1943+213: a candidate extreme BL Lac object}

\begin{figure}[tb]
\centering
\includegraphics[width=0.9\textwidth]{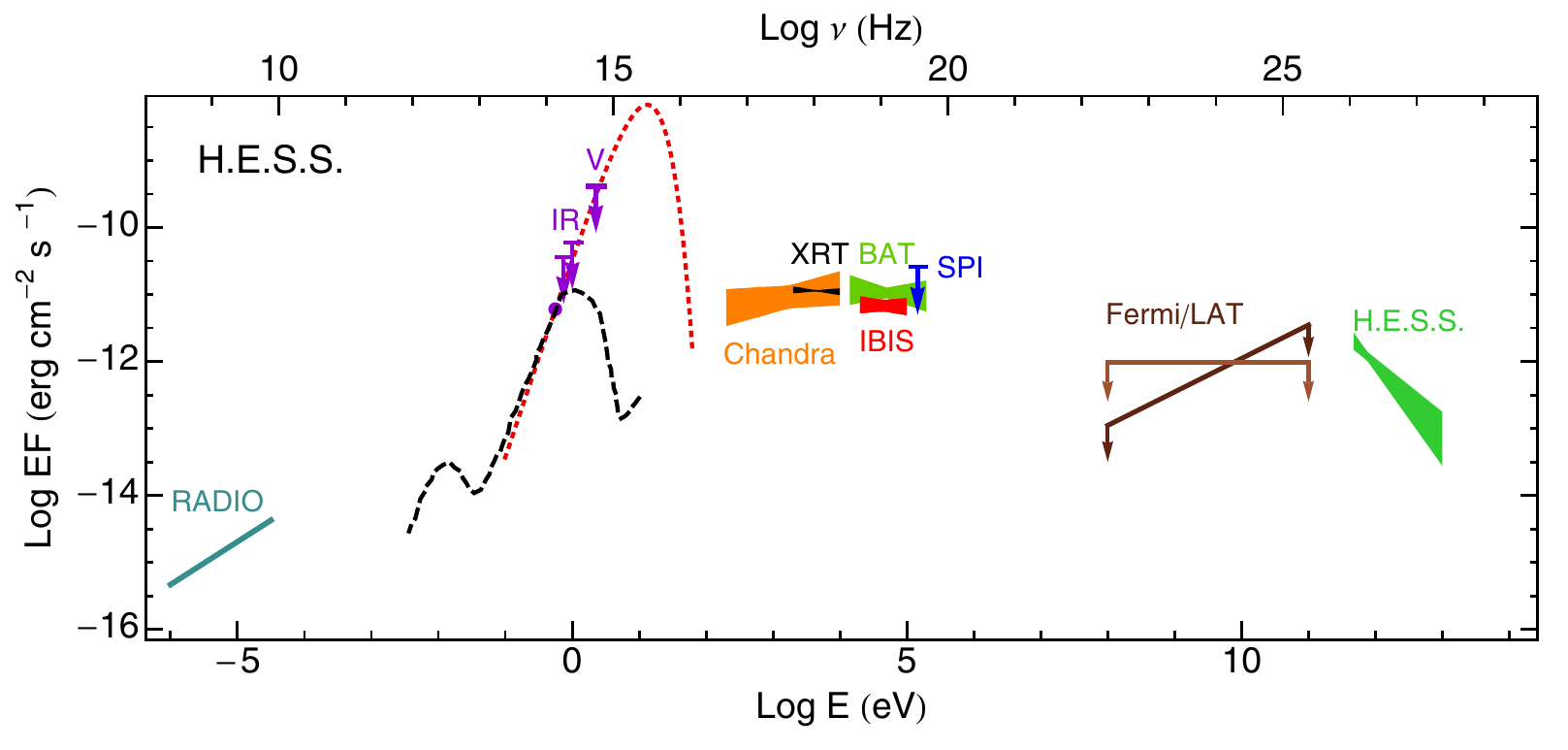}
\caption{Spectral energy distribution of  HESS\,J1943+213, a candidate extreme BL Lac object, and possible counterparts. The dashed black line is the spectral template of elliptical galaxies at a distance of 570 Mpc. The dotted (red) line is the blackbody spectrum of a massive star with parameters identical with the companion star in the gamma-ray binary LS 5039 at a distance of 25 kpc. The data in IR (2MASS) and visual (Swift/UVOT) bands were corrected for extinction. Further details can be found in \cite{abramowski:2011:hess:hessj1943:extremebllac}.\label{Fig:HESSj1943SED}}
\end{figure}

HESS~J1943+213 is a new point-like source detected at a significance level of $7.9 \sigma$ (post-trials) in the \He galactic plane survey. Details on the detection are reported in \cite{abramowski:2011:hess:hessj1943:extremebllac}. HESS~J1943+213 is located at
% Coordinates
RA(J2000) $=19^{\rm h} 43^{\rm m} 55^{\rm s} \pm 1^{\rm s}_{\rm stat} \pm 1^{\rm s}_{\rm sys}$,
DEC(J2000) $= +21^{\circ} 18' 8'' \pm 17''_{\rm stat} \pm 20''_{\rm sys}$. 
The source has a soft spectrum with photon index
% Photon index
$\Gamma = 3.1 \pm 0.3_{\rm stat} \pm 0.2_{\rm sys}$
% Photon flux
and a flux above 470 GeV of $(1.3 \pm 0.2_{\rm stat} \pm 0.3_{\rm sys}) \times 10^{-12}$ cm$^{-2}$~s$^{-1}$.
This source coincides with an unidentified hard X-ray source IGR~J19443+2117, which was proposed to have radio and infrared counterparts. There is no {\em Fermi}/LAT counterpart down to a flux limit of $6 \times 10^{-9}$ cm$^{-2}$~s$^{-1}$ in the 0.1--100 GeV energy range (95\% confidence upper limit calculated for an assumed power-law model with a photon index $\Gamma=2.0$). The data from radio to VHE gamma-rays do not show any significant variability.

The new \He, {\em Fermi}/LAT, and Nan\c{c}ay Radio Telescope observations were combined with pre-existing non-simultaneous multi-wavelength observations of IGR~J19443+2117 to investigate the likely source associations as well as the interpretation as an active galactic nucleus, a gamma-ray binary or a pulsar wind nebula (see Fig.~\ref{Fig:HESSj1943SED}).
% Conclusions
The lack of a massive stellar counterpart disfavors the binary hypothesis, while the soft VHE spectrum would be very unusual in case of a pulsar wind nebula. In addition, the distance estimates for Galactic counterparts places them outside of the Milky Way. All available observations favor an interpretation of HESS~J1943+213 as an extreme, high-frequency peaked BL Lac object with a redshift $z>0.14$, which would make it the first extragalactic source discovered serendipitously in the \He galactic plane survey.
 
%------------------------------------------------------------------------------------------------------------------------
% 1ES\,0414+009
%------------------------------------------------------------------------------------------------------------------------
\section{1ES\,0414+009}

\begin{figure}[tb]
\centering
\includegraphics[width=0.55\textwidth]{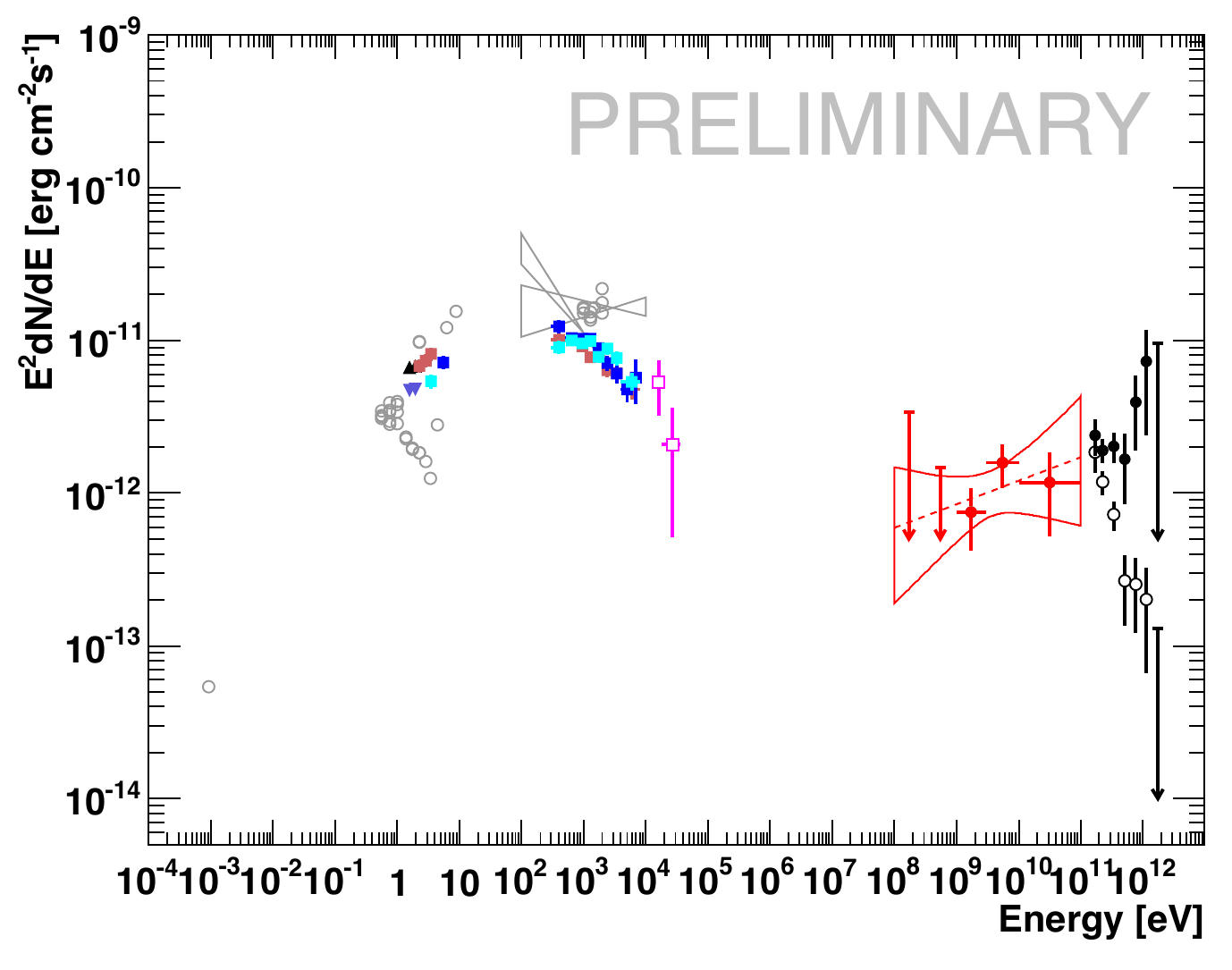}\hfill
\begin{minipage}[b]{0.4\textwidth}
\caption{Average SED of 1ES\,0414+009 with observations carried out between 2005 and 2009: \He (black fill/open circles with/without EBL correction respectively); \textit{Fermi}/LAT (red), \textit{Swift} BAT in 5 yrs (magenta), \textit{Swift} XRT \& UVOT (purple October 2006, blue January 2008 and cyan February 2008) and ATOM (triangles). (from \cite{volpe:2010a}) \label{Fig:1ES0414SED}}
\end{minipage}
\end{figure}

The BL Lac object 1ES\,0414+009 is located a redshift of $z = 0.287$ and harbors a super-massive black hole of mass  $2 \times 10^9\,$M$_\odot$. It has been identified as a high-frequency-peaked BL Lac (HBL) object with the synchrotron-emission peak located in the UV/soft-X-ray range.

\He observation of 1ES\,0414+009 have been performed between 2005 and 2009 for a total of 73.7\,hours, resulting in a detection with a statistical significance of 7.8\,standard deviations (details are presented in \cite{volpe:2010a}). The average energy spectrum measured by \He between 200\,GeV and 2\,TeV is well described by a power $dN/dE \propto E^{-\Gamma}$ with $\Gamma_{VHE} = 3.44 \pm 0.27_{stat} \pm 0.2_{sys}$ and an integral flux of $(1.83 \pm 0.21_{stat} \pm 0.37_{sys}) \times 10^{-12}$\,cm$^{-2}$s$^{-1}$ ($\sim 0.6$\% of the flux of the Crab Nebula) above an energy of 200\,GeV. No indication for flux or spectral variability was found in the \He data-set.
% Fermi
Observation with the Fermi/LAT (first 20 months of operations) result in a flux of $(2.3 \pm 0.2) \times 10^{-9}$\,erg\,cm$^{-2}$s$^{-1}$  between 200\,MeV and 10\,GeV and a spectrum well described by a power-law function with $\Gamma_{HE} = 1.85 \pm 0.18.$
% Swift
Swift/XRT observations show an X-ray flux of $F_{\rm2-10\,keV} = (0.8 - 1) \times 10^{-11 }$\,erg\,cm$^{-2}$s$^{-1}$, and a steep spectrum $\Gamma_x = 2.2 - 2.3$. Combining X-ray with Optical-UV data a fit with log-parabolic function locates the synchrotron peak around 0.1\, keV (see Fig.~\ref{Fig:1ES0414SED} for the SED).
%EBL
When corrected for EBL attenuation with a minimal EBL model \cite{franceschini:2008a} the VHE spectra smoothly extrapolates down to the Fermi spectrum, confirming previous constrains on the EBL density. The resulting hard intrinsic spectrum ($\Gamma \sim 1.9$), spanning from 0.2\,GeV to 2\,TeV, challenges simple one-zone SSC models.

%------------------------------------------------------------------------------------------------------------------------
% AP Lib
%------------------------------------------------------------------------------------------------------------------------
\section{AP Lib}

\begin{figure}[tb]
\centering
\includegraphics[width=0.9\textwidth]{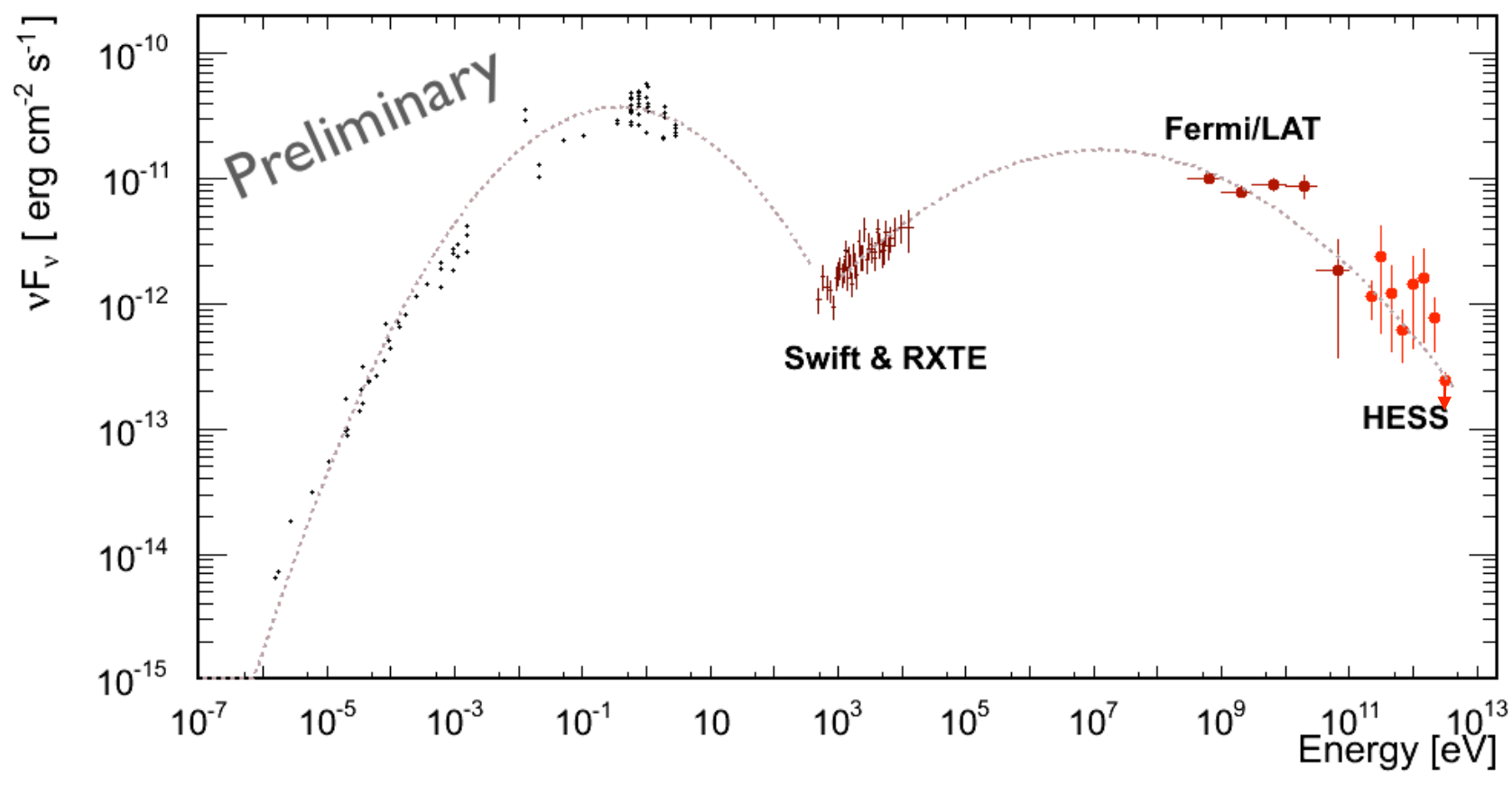}
\caption{SED of AP\,Librae with contemporaneous data from \textit{Swift}, Fermi/LAT, and H.E.S.S. Small black points in the radio to infrared are archival data from NED. Each component of the SED was fitted with with a 3rd degree polynomial in $\nu F_{\nu}$. (from \cite{fortin:2010a})\label{Fig:APLibSED}}
\end{figure}

The BL Lac type object AP\,Librae ($z=0.049$) was observed with \He for 11\,h of live time from May 11, 2010 to July 10, 2010 and was detected with a statistical significance of 7\,standard deviations (details can be found in \cite{fortin:2010a}). The energy spectrum is well described by a power law of index $\Gamma_{VHE}=2.5\pm0.2$. Following the detection, additional X-ray observations where performed with \textit{RXTE} and \textit{Swift}. The joint X-ray spectrum from $0.3\,{\rm
  keV}$ to $\sim 12\,{\rm keV} $ is well described by a power law ($P(\chi^2)=0.98$), a hard photon index of $\Gamma=1.58\pm0.06$ and a $2-10\,{\rm keV}$ flux of $F_{2-10\,{\rm keV}}=4.9_{-0.7}^{+0.8}\,{\rm erg}\,{\rm cm^{-2}}\,{\rm s^{-1}}$. This indicates that the second component of the spectral energy distribution, usually attributed to inverse-Compton radiation, is
  present at energies as low as ~0.3 keV. Fermi/LAT data on the source were taken from 4 August 2008 to 4 September
2010 (25 months). The HE energy spectrum is well described by a power law with spectral index of
$\Gamma=2.1\pm0.1$. No evidence for spectral curvature is found. The integral flux measured with Fermi/LAT between 300 MeV and 300 GeV is $F_{0.3-300\,{\rm GeV}}=(1.9\pm0.1) \times10^{-8}$\,cm$^{-2}$s$^{-1}$. When fitting the low and the high energy components with a 3rd degree polynomial in $\nu F_{\nu}$ synchrotron and inverse-Compton peak energies are found at $E_{s}^{\rm peak}\simeq 0.1\,{\rm eV}$ and $E_{IC}^{\rm peak}\simeq 10\,{\rm MeV}$, respectively. 

The SED of AP\,Librae is shown in Fig.~\ref{Fig:APLibSED}. One of the most striking features of the SED, compared to other VHE
emitting BL Lacs, is the width of the high-energy component, spanning over 10 decades in energy, extending from $\simeq 0.1\,{\rm keV}$ up to $\sim{\rm TeV}$ energies. Such wide distributions pose serious challenges for synchrotron self-Compton type models. 

%------------------------------------------------------------------------------------------------------------------------
% Mkn 421: 2010 VHE flare
%------------------------------------------------------------------------------------------------------------------------
\section{Mkn\,421: 2010 VHE flare}

\begin{figure}[tb]
\centering
\includegraphics[width=0.6\textwidth]{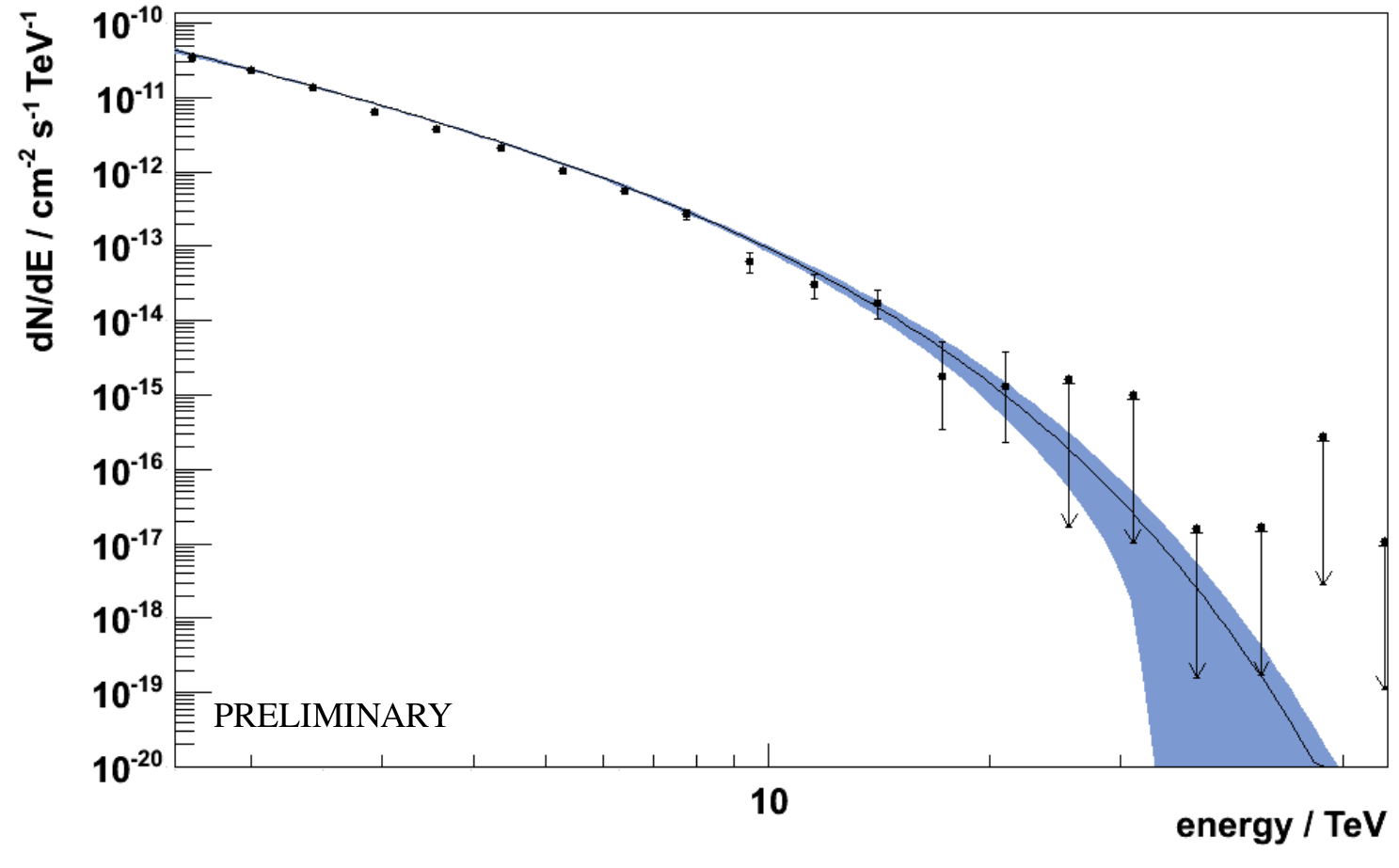}\hfill
\begin{minipage}[b]{0.35\textwidth}
\caption{Differential energy spectrum of Mkn\,421 as measured by H.E.S.S. in Feb. 2010 in 5.4\,h of live-time. The spectral shape can be parameterized by $dN/dE = \phi_0 (E/E_0)^{-\Gamma} \mathrm{exp}(-E/E_{cut})$,
with a flux normalization $\phi_0 = (1.96\,\pm\,0.32)\,10^{-11}$\,cm$^{-2}$s$^{-1}$TeV$^{-1}$,
a spectral index $\Gamma = 2.05\,\pm\,0.22$, a cutoff energy $E_{cut} = (3.4\,\pm\,0.6)$\,TeV,
and a decorrelation energy $E_0 = 2.739$\,TeV.
(from \cite{tluczykont:2010a})\label{Fig:Mkn421}}
\end{minipage}
\end{figure}

The high-frequency peaked BL Lac object Mkn\,421 ($z=0.03$) showed a strong outburst of activity in February 2010, first reported by the VERITAS collaboration (see A. Furniss, this proceedings). Following the trigger, H.E.S.S. observed Mkn\,421 for a total of 5.4\,h of good quality observation time from Feb. 17 to 20, 2010, at an average zenith angle of 62.4$^\circ$  (see \cite{tluczykont:2010a} for details). A total of 2112 excess events (2188 on-events and 838/11 off-events) were detected yielding a significance of 86.5 standard deviations (s.d.). Significant nightly variability has been detected in the data-set. The time averaged energy spectrum is well described by a power law function with exponential cut-off  $dN/dE = \phi_0 (E/E_0)^{-\Gamma} \mathrm{exp}(-E/E_{cut})$, with flux normalization $\phi_0 = (1.96\,\pm\,0.32)\,10^{-11}$\,cm$^{-2}$s$^{-1}$TeV$^{-1}$, a spectral index $\Gamma = 2.05\,\pm\,0.22$, cutoff energy $E_{cut} = (3.4\,\pm\,0.6)$\,TeV, and de-correlation energy $E_0 = 2.739$\,TeV (see Fig.~\ref{Fig:Mkn421}). The last significant energy bin ($>3$\,s.d.) in the spectrum extends beyond 20\,TeV (highest energy observed from this source), making these observations relevant for studies of the density of the mid-infrared of the extragalactic background light (EBL). The measured spectral shape is compatible with previous results, implying a stability of the central engine over the epochs.

%------------------------------------------------------------------------------------------------------------------------
% M\,87: 2010 VHE campaign
%------------------------------------------------------------------------------------------------------------------------
\section{M\,87: 2010 VHE campaign}
The giant radio galaxy \M with its proximity (16\,Mpc), famous jet, and its very massive black hole ($3-6 \times10^9$M$_\odot$) provides a unique laboratory to investigate the origin of VHE $\gamma$-ray emission from relativistic outflows and the vicinity of super-massive black holes (SMBH). \M has been established as a VHE emitter since 2005 \cite{aharonian:2003b,aharonian:2006:hess:m87:science}. The VHE emission displays strong variability on time-scales as short as a day \cite{aharonian:2006:hess:m87:science}. The detection of a VHE outburst contemporaneous with a flare of the radio core, directly imaged by 43\,GHz VLBA observations with sub-parsec resolution, lead to the interpretation that the VHE emission is likely generated in the close vicinity of SMBH \cite{acciari:2009b:m87joinedcampaign:science}. To further investigate this exciting possibility, VHE monitoring campaigns on the source have been performed, with the results of the latest one, lead by MAGIC and VERITAS in 2010, being reported in this proceedings. A more complex MWL picture is found, which eludes an easy interpretation. Details can be found in M. Raue et al. in this proceedings.

%------------------------------------------------------------------------------------------------------------------------
% PKS\,2155-304: Variability studies and monitoring
%------------------------------------------------------------------------------------------------------------------------
\section{PKS\,2155-304: Variability studies and monitoring}

PKS\,2155-304 is a well-known high-frequency-peaked BL\,Lac object with VHE emission. It displayed an extreme flux outburst in 2006, with variability time-scales down to a few minutes \cite{aharonian:2007:hess:pks2155:bigflare}. To further investigate the spectral and temporal variability of the VHE emission in the source a large \He data-set ($\sim90$\,h), collected between 2005 to 2007, has been analyzed for its timing and spectral properties \cite{abramowski:2010:hess:pks2155:timing}. The source was found in low flux state from 2005 to 2007 with the exception of a set of exceptional flares  occurring in July 2006. The quiescent state can be characterized by a mean flux  level of $(4.32\pm0.09_\mathrm{stat}\pm0.86_\mathrm{syst}) \times 10^{-11}\,{\rm cm^{-2}}\,{\rm s^{-1}}$ above $200\,{\rm GeV}$, or approximately $15\%$ of the Crab Nebula, and a power law energy spectrum with photon index of $\Gamma=3.53\pm0.06_\mathrm{stat}\pm0.10_\mathrm{syst}$. During the flares doubling timescales of $\sim 2\,{\rm min}$ are found \cite{aharonian:2007:hess:pks2155:bigflare}. The variation of the photon index with flux was investigated over two orders of magnitude in flux, yielding different behavior at low and high fluxes, which is a new phenomenon in the VHE domain for $\gamma$-ray emitting blazars. The variability amplitude, characterized by the fractional r.m.s. $F_{\rm var}$, is strongly energy-dependent ($\propto E^{0.19\pm0.01}$). The light curve r.m.s. correlates with the flux. This is the signature of a multiplicative process which can be accounted for as a red noise with a Fourier index of $\sim 2$.

To further investigate the long-term variability of the source at VHE (H.E.S.S.), HE (\textit{Fermi}/LAT), and optical (ATOM, ROTSE), H.E.S.S. is performing a long-term monitoring campaign on PKS\,2155-304 in 2010 and 2011 with observations every 2nd night all throughout the H.E.S.S. observability window (June/July to December; gaps due to moonlight).

%------------------------------------------------------------------------------------------------------------------------
% Conclusion \& Outlook
%------------------------------------------------------------------------------------------------------------------------
\section{Conclusion \& Outlook}

The H.E.S.S. extragalactic science program is still going strong. Exciting discoveries have been made in the past two years (HESS\,J1943+213 \cite{abramowski:2011:hess:hessj1943:extremebllac}, AP\,Librae \cite{fortin:2010a}, 1ES\,0414+009 \cite{volpe:2010a}, SHBL\,J001355.9-185406, and 1RXS\,J101015.9-311909 \cite{becherini:2010a}). Known sources keep to deliver compelling science \cite{abramowski:2010:hess:pks2155:timing,acciari:2009b:m87joinedcampaign:science,tluczykont:2010a}.

% H.E.S.S. II
In 2012 the H.E.S.S. II telescope will start operating extending the H.E.S.S. energy range down to $\sim$20-30\,GeV and increasing the overall sensitivity by a factor $\sim$2. H.E.S.S. II will enable to extend AGN studies to higher redshifts and to probe the transition region between the Fermi/LAT and the VHE energy range with high sensitivity.

% CTA
The Cherenkov Telescope Array (CTA) \cite{cta:2010:conceptionaldesignreport}, which is expected to start construction in 2014, will deliver a factor of 10 improvement in sensitivity over current generation instruments and an extended energy range from 20\,GeV to $\sim$100\,TeV. Two sites are planned, with the northern side being dedicated to extra-galactic science and low energy thresholds.

%------------------------------------------------------------------------------------------------------------------------
% Acknowledgements
%------------------------------------------------------------------------------------------------------------------------
 \ack
 % LEXI
 The author acknowledges support by the LEXI program of the state of Hamburg, Germany.
% HESS
--- The H.E.S.S. Collaboration acknowledges support of the Namibian authorities and of the University of Namibia
in facilitating the construction and operation of H.E.S.S., as is the support by the German Ministry for Education and
Research (BMBF), the Max Planck Society, the French Ministry for Research,
the CNRS-IN2P3 and the Astroparticle Interdisciplinary Programme of the
CNRS, the U.K. Science and Technology Facilities Council (STFC),
the IPNP of the Charles University, the Polish Ministry of Science and 
Higher Education, the South African Department of
Science and Technology and National Research Foundation, and by the
University of Namibia. We appreciate the excellent work of the technical
support staff in Berlin, Durham, Hamburg, Heidelberg, Palaiseau, Paris,
Saclay, and in Namibia in the construction and operation of the
equipment.
% NASA ADS
--- This research has made use of NASA's Astrophysics Data System.

%------------------------------------------------------------------------------------------------------------------------
% Bibliography
%------------------------------------------------------------------------------------------------------------------------

\section*{References}

\def\Journal#1#2#3#4{{#4}, {#1}, {#2}, #3}
\def\NAT{Nature}
\def\AAA{A\&A}
\def\ApJ{ApJ}
\def\AJ{Astronom. Journal}
\def\Aph{Astropart. Phys.}
\def\ApJS{ApJSS}
\def\MNRAS{MNRAS}
\def\NIM{Nucl. Instrum. Methods}
\def\NIMA{Nucl. Instrum. Methods A}
% Bibliography and bibfile
\def\aj{AJ}%
          % Astronomical Journal
\def\actaa{Acta Astron.}%
          % Acta Astronomica
\def\araa{ARA\&A}%
          % Annual Review of Astron and Astrophys
\def\apj{ApJ}%
          % Astrophysical Journal
\def\apjl{ApJ}%
          % Astrophysical Journal, Letters
\def\apjs{ApJS}%
          % Astrophysical Journal, Supplement
\def\ao{Appl.~Opt.}%
          % Applied Optics
\def\apss{Ap\&SS}%
          % Astrophysics and Space Science
\def\aap{A\&A}%
          % Astronomy and Astrophysics
\def\aapr{A\&A~Rev.}%
          % Astronomy and Astrophysics Reviews
\def\aaps{A\&AS}%
          % Astronomy and Astrophysics, Supplement
\def\azh{AZh}%
          % Astronomicheskii Zhurnal
\def\baas{BAAS}%
          % Bulletin of the AAS
\def\bac{Bull. astr. Inst. Czechosl.}%
          % Bulletin of the Astronomical Institutes of Czechoslovakia 
\def\caa{Chinese Astron. Astrophys.}%
          % Chinese Astronomy and Astrophysics
\def\cjaa{Chinese J. Astron. Astrophys.}%
          % Chinese Journal of Astronomy and Astrophysics
\def\icarus{Icarus}%
          % Icarus
\def\jcap{J. Cosmology Astropart. Phys.}%
          % Journal of Cosmology and Astroparticle Physics
\def\jrasc{JRASC}%
          % Journal of the RAS of Canada
\def\mnras{MNRAS}%
          % Monthly Notices of the RAS
\def\memras{MmRAS}%
          % Memoirs of the RAS
\def\na{New A}%
          % New Astronomy
\def\nar{New A Rev.}%
          % New Astronomy Review
\def\pasa{PASA}%
          % Publications of the Astron. Soc. of Australia
\def\pra{Phys.~Rev.~A}%
          % Physical Review A: General Physics
\def\prb{Phys.~Rev.~B}%
          % Physical Review B: Solid State
\def\prc{Phys.~Rev.~C}%
          % Physical Review C
\def\prd{Phys.~Rev.~D}%
          % Physical Review D
\def\pre{Phys.~Rev.~E}%
          % Physical Review E
\def\prl{Phys.~Rev.~Lett.}%
          % Physical Review Letters
\def\pasp{PASP}%
          % Publications of the ASP
\def\pasj{PASJ}%
          % Publications of the ASJ
\def\qjras{QJRAS}%
          % Quarterly Journal of the RAS
\def\rmxaa{Rev. Mexicana Astron. Astrofis.}%
          % Revista Mexicana de Astronomia y Astrofisica
\def\skytel{S\&T}%
          % Sky and Telescope
\def\solphys{Sol.~Phys.}%
          % Solar Physics
\def\sovast{Soviet~Ast.}%
          % Soviet Astronomy
\def\ssr{Space~Sci.~Rev.}%
          % Space Science Reviews
\def\zap{ZAp}%
          % Zeitschrift fuer Astrophysik
\def\nat{Nature}%
          % Nature
\def\iaucirc{IAU~Circ.}%
          % IAU Cirulars
\def\aplett{Astrophys.~Lett.}%
          % Astrophysics Letters
\def\apspr{Astrophys.~Space~Phys.~Res.}%
          % Astrophysics Space Physics Research
\def\bain{Bull.~Astron.~Inst.~Netherlands}%
          % Bulletin Astronomical Institute of the Netherlands
\def\fcp{Fund.~Cosmic~Phys.}%
          % Fundamental Cosmic Physics
\def\gca{Geochim.~Cosmochim.~Acta}%
          % Geochimica Cosmochimica Acta
\def\grl{Geophys.~Res.~Lett.}%
          % Geophysics Research Letters
\def\jcp{J.~Chem.~Phys.}%
          % Journal of Chemical Physics
\def\jgr{J.~Geophys.~Res.}%
          % Journal of Geophysics Research
\def\jqsrt{J.~Quant.~Spec.~Radiat.~Transf.}%
          % Journal of Quantitiative Spectroscopy and Radiative Trasfer
\def\memsai{Mem.~Soc.~Astron.~Italiana}%
          % Mem. Societa Astronomica Italiana
\def\nphysa{Nucl.~Phys.~A}%
          % Nuclear Physics A
\def\physrep{Phys.~Rep.}%
          % Physics Reports
\def\physscr{Phys.~Scr}%
          % Physica Scripta
\def\planss{Planet.~Space~Sci.}%
          % Planetary Space Science
\def\procspie{Proc.~SPIE}%
          % Proceedings of the SPIE
          
\providecommand{\newblock}{}

\end{document}